\documentstyle[preprint,aps,prl]{revtex}
\begin{document}
\title{The Dynamics of a Meandering River 
}
\author{
${\mbox{T.\,B.\ Liverpool},}$\cite{Corres,TBLpresAdd}
and
${\mbox{S.\,F.\ Edwards}}$
}
\address{
Cavendish Laboratory, University of Cambridge,\\
Madingley Road, Cambridge CB3 0HE, U.K.}
\date{\today}
\maketitle
\begin{abstract}
We present a statistical model of a meandering river
on an alluvial plane which is
motivated by the physical non-linear
dynamics of the river channel
migration and by describing heterogeneity of the terrain by 
noise. We study the dynamics analytically and numerically. 
The motion of the river channel
is unstable and we show that by inclusion of
the formation of ox-bow lakes, the system 
may be stabilised. We then calculate the steady state and 
show that it is in agreement with simulations and 
measurements of field data.
\\
\\
\end{abstract}

Meandering rivers are ubiquitous in nature 
and can be found all over the world.
It is a signature of their intellectual challenge that they
have captured the imagination of scientists in
quite differing fields~\cite{Einstein26}.
They are important not only as examples of a system far from equilibrium 
undergoing an interesting dynamics but 
because abandoned channels of the main river become silted up and over
thousands of years these silted channels become compressed to
form shale beds between which oil is often found. 
Previous models that have tried to study river meanders fall into
two broad classes. The first could be considered as {\em empiric} 
and are justified by the fact that 
Leopold and Wolman~\cite{LeoWol60}
showed by an analysis of field data from around the world, that there
was a nearly constant ratio of
radius of curvature to meander length and of radius of
curvature to meander width. This `explains' the visual
{\em `scale invariance'} when one observes rivers. They therefore
concluded that `` meander geometry is related in some {\em unknown
manner} to a more general mechanical theory''. Consequently there have 
since been a number of papers~\cite{LangLeo66,Ferg76,HoKn84} 
that modelled meanders as variants of random 
walks with essentially no justification. The second type is 
characterised by an analysis of the complex flow inside the 
channel~\cite{Call69,Enge73,Ikeda1,ParDA83}
and as a result yields relatively little information about large 
scale or long time features of the river pattern. 
In this letter, 
a slightly different approach is taken; we attempt to find the
simplest possible model of the behaviour of a meandering
river which still retains the essential physics. 
The model is statistical in the sense that there 
is a spatio-temporal random aspect to the dynamics. 
Statistical models have been introduced in the study of 
river networks~\cite{KraMa91,LeNag93,Rin93}
but have not been used in the study of single channels.
Also, to the authors knowledge,
a stable state has not been constructed for any dynamic models of 
river meanders.

We consider the river on the length scale such that its 
channel may be considered to be a curve of length 
$L$ where $L\rightarrow \infty$,  
in two dimensions, ${\bf r}(s,t)=[x(s,t),y(s,t)]$ 
where $s$ refers to the arc length position on the curve. We include
all the major features of the dynamics of the channel and using them  
calculate the steady state. The derivation is outlined here and details 
will be reported elsewhere~\cite{tannie}.  
Our theory of meanders therefore acts as a   
connection between both previous approaches.  

To calculate the equation of motion (EOM) of the river 'curve', 
we must begin by an analysis~\cite{Enge73,Ikeda1} of 
the flow of water in a curved channel. The channel is considered to be 
narrow with width $b$ and 
we can describe the coordinates of the channel as $s$ and $n$ corresponding
to moving along the centre-line of the channel and perpendicular to it,  
and note $n=n(s)$. 
We transform the Navier-Stokes 
equation for incompressible flow to the curved channel coordinates system. 
Using the similarly transformed continuity equation and assuming no stress 
at the top surface and bed stresses at the bottom of the
channel, it can be shown using simple 
arguments~\cite{Ikeda1} that the major effect of this is that 
a velocity gradient develops across the channel with the
velocity of water at the outer bank being greatest. That is if the outer bank 
corresponds to $n=b/2$ and the inner bank $n=-b/2$, then the velocity of 
fluid in the channel is given 
to leading order by $u(n) \approx u_0(1 + n~ \kappa(s))$ where $\kappa(s) = 
|{\partial^2 {\bf r}}/{\partial s^2}|$ is the local curvature and 
$u_0$ is a constant. We do not include gravity so the 
plane is flat but we do have a pressure gradient driving the flow 
through the channel.
In the analysis above the approximation is made that 
the channel is locally smooth, 
i.e there are no sharp changes 
in direction meaning the local curvature $\kappa(s)$ is small. 
Assuming erosion on the outer bank is balanced by deposition 
on the inner bank and that the rate of erosion on the outer bank is 
proportional to the velocity, the rate of normal, i.e. 
perpendicular to the curve, migration of 
the channel can be described by the relationship 
\begin{equation}
R^0_m \propto \kappa(s) \, .
\end{equation}
The condition that the 
curve is locally smooth is satisfied by adding another piece to the
rate of migration; i.e. 
\begin{equation}
R^1_m \propto  - \kappa^3 + \frac{\partial^2\kappa}{\partial s^2} \, .
\end{equation}
This is obtained from the Rayleighian dynamics~\cite{EdwFre74} 
of a 'global' free energy 
which minimises curvature 
along the whole meandering river curve, 
${\cal F}[r]=\int ds |{\partial^2 {\bf r}}/{\partial s^2}|^2$. 
Evolution of curves undergoing similar dynamics as these first 
two bits of $R_m$ have 
been studied 
in the context of crystal growth~\cite{BroKKL1,BroKKl2} 
and chemical front motion~\cite{PeGol94}. 
Finally, we  have a dynamic noise term to model heterogeneities in
space, such as different types of rock, and
in time, for example, 
 variation in flow of the river and re-eroding parts of
the plain that have been eroded previously.
\begin{equation}
R^2_m \propto \eta(s,t) \, .
\end{equation}
Dynamic noise is more appropriate than quenched noise because 
the river is constantly reworking its plain as it meanders and over 
long periods changing the nature of the terrain.
For simplicity we take Gaussian uncorrelated noise. 

To get the correct EOM we must appeal to reparametrisation invariance
because the arc length, $s$ is not constant in time. 
Physically, it is known that the normal velocity of a curve, in this case 
$\hat{\bf n} \cdot \partial_t {\bf r}(s,t) = \sum_{i=0}^2 a_i R^i_m$\, , 
($\hat{\bf n}$ 
is the unit normal) is 
independent of its parametrisation so for simplicity 
the EOM may be considered first in the 
'normal gauge' where the tangential velocity of the curve is zero.
Then using the Frenet formulae from differential geometry the EOM in the 
'arc length gauge' may then be calculated to give,
\begin{equation}
\frac{\partial}{\partial t} {\bf r}(s,t) + \frac{\partial{\bf r}}{\partial s}
\int^s ds'\kappa(s')V = \hat{\bf n}V
\label{eq:dtr}
\end{equation}
where  the normal velocity may be written 
$V = A \kappa(s) -B \left( \kappa^3 - {\partial^2 \kappa}/{\partial s^2}
\right) + \eta(s,t)$  
$  
= -\hat{\bf n} \cdot \left( A {\partial^2{\bf r}}/{\partial s^2} 
+B {\partial^4{\bf r}}/{\partial s^4} \right) + \eta(s,t) \, .
$
The smallest length-scale of the system is therefore 
$\sqrt{B/A}$. 
Recalling that $\partial {\bf r}/\partial s = {\bf \hat{s}}$, 
the unit tangent vector to the curve, one sees that the second term on the 
lhs of equation(\ref{eq:dtr})
is a non-local tangential velocity 
corresponding to a 'stretching' or lengthening of the curve due the 
normal growth. To calculate the effective motion of the 
curve we replace the tangential velocity by its average, i.e. 
because we are dealing with a Langevin equation with the
prescence of a noise term, and we assume the system will
always be near some steady state, we make the 
{\it pre-averaging} approximation where complicated parameters
in the equation
are replaced by their mean values 
, so we define $\zeta(A,B) = 
\left\langle \int^s \kappa V ds'\right\rangle =  
\int {\cal D}{\bf r} \int^s \kappa V ds' P([{\bf r}],t)$, 
and consequently we may write 
the EOM of the river as
\begin{equation}
\frac{\partial}{\partial t} {\bf r}(s,t) + 
\zeta \frac{\partial{\bf r}}{\partial s} = 
- A \frac{\partial^2{\bf r}}{\partial s^2} - 
B \frac{\partial^4{\bf r}}{\partial s^4}  + \eta(s,t) \, .
\label{eq:rivdyn}
\end{equation}
where
\begin{equation}
\langle\eta(s,t)\rangle = 0 \ , \ \langle \eta(s,t)
\eta(s',t') \rangle = D \delta(s-s')\delta(t-t')
\nonumber
\end{equation}

We have thus calculated the first part of the problem, 
the {\it local} dynamics of the meandering river as a 
non-linear noisy equation. Noisy non-linear equations have been 
widely studied in the context of surface deposition~\cite{HaZh95} 
particularly when they lead to critical behaviour. 
Unlike these models meandering river systems {\it do evolve} into 
statistically invariant stable states.
This equation also bears some superficial similarity to the Rouse
equation of polymer dynamics~\cite{DoiEds}, but its behaviour is actually
quite different and is in fact extremely unstable and would
lead to proliferation of length. This can be seen by 
a transformation to Fourier modes 
${\bf r}_k = \int ds {\bf r}(s)\exp\{iks\}$ gives  
${\bf r}_k(t)= {\bf r}_k(0)\exp\{\Delta(k)t - ik\zeta t\}$ with 
$\Delta(k)=Ak^2-Bk^4$
so there is always a band of unstable modes for which $\Delta>0$.

In the form of equation(\ref{eq:rivdyn}) above, 
the essential dynamics is physically transparent and numerical 
interpretation very efficient. Similar equations (but without noise) 
such as 
those found in viscous fingering~\cite{Ben86} 
could also be studied in this way. 

We numerically integrated equation(\ref{eq:rivdyn})
above starting from an initial condition of a straight line and 
from Figure 1 it can be seen that meandering patterns are developed.
Further evolution of the pattern leads to a plane filling, many time
self-intersecting 
curve~\cite{tannie}.

We now include the {\it non-local} 
part of the dynamics, the formation of ox-bow lakes;
when the river intersects
itself it loses that part of the curve between
the two points of contact. This is the way the system is stabilised.
We stress that locality here refers to $s$ the position along the
river and not ${\bf r}$ the position in space.

Before we proceed, it is useful to rewrite the EOM as a Fokker-Planck 
equation~\cite{vanKam} so that the probability distribution 
$P([{\bf r}(s)],t)$ of 
the points on the river obey the equation
\begin{equation}
\frac{\partial}{\partial t}P =  {\cal O}_r P 
\end{equation}
where 
\begin{displaymath}
 {\cal O}_r = \int dL \int^L_0 ds 
\frac{\delta}{\delta {\bf r}(s)} \cdot \left(
\frac{\delta}{\delta {\bf r}(s)} + \zeta~\frac{\partial{\bf r}}{\partial s} 
+ A~\frac{\partial^2{\bf r}}{\partial s^2} + 
B~\frac{\partial^4{\bf r}}{\partial s^4} \right) 
\end{displaymath}
The integral over $L$ acknowledges the fact that the length does not remain 
constant. Since the river is taken as very long,
the fact that the length $L$ varies is not so important as we assume
that  near the steady state
$\left\langle L \right\rangle$ is a constant in time.

The oxbow interaction can be considered as a transition
between states and we can define a transition probability in
a Boltzmann type master
equation~\cite{vanKam}. The equation will
be of the form
${\partial_t}P_a =
{\cal O}_r P_a + \sum_b \left[ T_{ab} P_b - T_{ba}P_a
\right]
$ where $P_b$ refers to states scattering to $P_a$
or being scattered to from $P_a$
via ox-bow creation and $T_{ab}$ is the
{\em `transition matrix'}.
For our purposes, 
it is sufficient to make the association
$P([{\bf r}(s)],t)=\lim_{N\rightarrow\infty}P([{\bf r}_1,\ldots,{\bf r}_N],t)$
so the functional can be taken as the infinite limit of a discrete function 
of many variables. In this notation the creation of an ox-bow happens when 
the curve 
intersects itself at $i$ and $j$ where without loss of generality,
 $j>i$ 
corresponds to the transformation of an initial 
probability 
distribution $P_I$ for which ${\bf r}_i={\bf r}_j$ 
to a final distribution $P_F$ 
for which the portion of the curve between 
${\bf r}_i$ and ${\bf r}_j$ has vanished 
and so the curve must be {\it re-labelled} such that 
$\forall \, k>j,\, {\bf r}_k\rightarrow {\bf r}_{k-(j-i)}$. 
We can obtain an approximation of 
the final distribution in terms of the initial 
distribution by a generalisation of a Taylor expansion to 
functionals so that in the continuum notation, the initial and final 
distributions are given by, respectively, 
$P_I[{\bf r}(s),L,t]$ and 
$P_F[{\bf r}(s),L-\delta L, t+\delta t]$ where $\delta L$ is the length
of the ox-bow.
Therefore the creation of an ox-bow by intersection at 
$s_i \ \ {\rm and} \ \  s_j$ with $s_i < s_j$ is described by
\begin{eqnarray}
& & P_F[\{{\bf r}(s)\},s_i,s_j]- P_I[\{{\bf r}(s)\},s_i,s_j] \approx 
\nonumber \\ 
& & \int dL \int_{s_i}^{L} ds'\frac{\delta P_I}{\delta {\bf r}(s')}
\left({s_{ij}}\frac{\partial {\bf r}}{\partial s'} + \frac{s_{ij}^2}{2}
\frac{\partial^2{\bf r}}{\partial s'^2} + O\left({s_{ij}^3}\right)
\right)
\label{eq:pfpi}
\end{eqnarray} where $s_{ij}=|s_i-s_j|$. 
Now this will only work if we can consider $|s_i-s_j|$
to be small and so by default
this approximation will only be able to deal with small loops, 
i.e. $|s_i-s_j|<<L$.

The transition probability $\tau$ will depend on the rate of approach of 
points ${\bf r}(s_i)$ and ${\bf r}(s_j)$ and so 
$
\tau = \lambda \partial_t\left\{
\int \int ds_i ds_j 
\delta[{\bf r}(s_i) - {\bf r}(s_j)]\right\} 
$
 where $\lambda$ is a constant.
The sum over $b$ in the Fokker-Planck equation corresponds to 
a configurational sum over the state being scattered to or from.

To determine the effect of ox-bows, we make the assumption that their effect 
added to our prescription 
will 
give us, in the statistical sense, a steady state. In a sense, 
we construct such a steady state.
The EOM becomes  
$\partial_t P_s = {\cal O}_r P_s + {\cal I}_{ox}P_s$  where 
${\cal I}_{ox}$ is an effective interaction due to the ox-bows.  
Because we are in the steady state,  
$\left\langle \partial_t P_s 
\right\rangle = 0$ so that we may write 
$\left\langle P_s\right\rangle \propto \exp\left\{-F[{\bf r}]\right\}$ where 
$F=F_0[{\bf r}]+F'[{\bf r}]$ and the operator acting on $P_s$ is given by
 ${\cal I}_{ox}P_s = 
\int ds {\delta}/{\delta {\bf r}}({\delta F'}/{\delta {\bf r}}P_s)$ 
and  the other part, ${\delta F_0}/{\delta {\bf r}(s)}=  
+ \zeta {\partial{\bf r}}/{\partial s}
+ A {\partial^2{\bf r}}/{\partial s^2} +
B {\partial^4{\bf r}}/{\partial s^4}$. \linebreak 
$F_0$ is the part of the free energy from the local dynamics and we want 
$F'$ the part from the ox-bow to somehow cancel out the instabilities 
in the local EOM. It turns out that $F'$ has two parts.

The first part of the interaction can be dealt by looking at 
the average value of the transition away from the present state by 
loss of an ox-bow, and we find  
an interaction term 
\begin{equation}
F_1' \simeq \lambda \int ds ds'
\delta[{\bf r}(s)-{\bf r}(s')] \, .
\end{equation}
To obtain this  we used 
the fact that $\left\langle \partial_t {\bf r} 
\right\rangle = {\delta F}/{\delta {\bf r}(s)}$ and 
$\delta \left\langle P_s \right\rangle / \delta {\bf r} = 
\left\langle P_s \right\rangle \delta F / \delta {\bf r} $
and we rewrote the transition probability in the form
$\tau=\lambda \partial_t {\bf r}(s_i) \cdot
{\delta}/{\delta {\bf r}(s_i)} \left\{ \int \int ds_i ds_j
 \delta[{\bf r}(s_i) - {\bf r}(s_j)]\right\}
$. This is the same form of term that would be obtained from 
the dynamics of the self-avoiding walk (SAW)~\cite{Edw66} Hamiltonian.

The second part is obtained from the transition into the present state from 
another by loss of an ox-bow. Here we find that the effect of the transition 
is given by 
\begin{equation}
\frac{\delta}{\delta {\bf r}} \cdot \frac{\delta F_2'}{\delta {\bf r}}P 
\simeq \frac{\delta P}{\delta {\bf r}}
\left(g_1(\lambda) \frac{\partial {\bf r}}{\partial s} +
g_2(\lambda)
\frac{\partial^2{\bf r}}{\partial s^2} + \ldots \right)
\end{equation}
where $g_{\alpha}(\lambda)=\left\langle \lambda \int ds_ids_j\partial_t
\{\delta[{\bf r}(s_i)-{\bf r}(s_j)]\}|s_i-s_j|^{\alpha} \right\rangle$ are a simple function of $\lambda$ and measures 
the average 
rate of approach of points along the curve.
The effect of this will be that
if we consider the original equation we have a renormalisation
of parameters due to the inclusion of the effective field of the
ox-bow interaction, 
$A \rightarrow \tilde{A}  =  A - g_2(\lambda)$ and
$\zeta \rightarrow \tilde{\zeta}  =  \zeta - g_1(\lambda)$.
For the system to remain in a stable state, we
{\it must} have $A-g_2(\lambda)= -2AD/B$ and $\zeta=g_1(\lambda)$. 
We get the same results from independent scaling arguments. 
Combining both parts, we obtain a stable state that should correspond 
to a system always near that of a SAW confirming 
recent lattice simulations and measurements from field data~\cite{LBE95}.

We also included the creation of ox-bows to the numeric analysis of the
meander dynamics by a relabelling procedure on contact,
 and we generated stable meander configurations
in contrast to the situation when they are not present. In Figure 
2 we have the configurations after $200,000$ time-steps for 
the system with and without the inclusion of the ox-bow mechanism 
starting from an initial configuration of a straight line in both 
cases. The mean spread of the river path $Y_g=
\sqrt{\langle (Y-\bar{Y})^2 \rangle}$
as a function of length $L$
was also
calculated and we obtained the relationship
$Y_g \sim L$ for short distances
and a cross-over to
$Y_g \sim L^{1/2}$ for
long distance, agreeing with
field data~\cite{LBE95}.

We now briefly consider some extensions to the model to make it 
more realistic; these will be discussed fully elsewhere~\cite{tannie}.  
In the expression for the fluid velocity for the 
channel, higher order terms are of the type $\delta u \sim \int^s  
\kappa \exp\{-\gamma(s-s')\}$ where $\gamma$ is a constant depending 
on the parameters of the flow and infract it can be shown~\cite{tannie}
that such corrections lead to an addition of a constant vector 
to the EOM. In other words it gives the river a well defined direction 
and does not change the dynamics described above. 
 It is also instructive to consider higher order 
terms in the Taylor expansion undertaken to describe the ox-bow 
creation. These terms lead to a renormalisation~\cite{tannie} of the noise 
so that  
$D \rightarrow \tilde{D}$. Finally we consider the effect of 
'valley confinement' and slope to the river channel. These will result 
in a potential $U[{\bf r}]$ of the form $(g,0) + (0,hy)$ added to the EOM.

In conclusion, we have developed a phenomenological model of river 
meanders which reproduces all the major features seen in nature 
motivated by an analysis of flow in a meandering channel. An essential 
feature of the model is the competition of ox-bow creation with a curvature 
instability of the river channel. It is 
particularly interesting that with the inclusion of ox-bows to 
the motion of the river, stable meanders form for {\it all} values of 
$A,B,D$. The ratio $B/A$ gives the lower length-scale of the 
river pattern or the size of the smallest ox-bows and the value 
of $D$ gives the 'time scale' over which the system retains memory
of its previous behaviour. This explains the visual scale 
invariance of meandering river systems.

This work was funded by EPSRC.
It is a pleasure to acknowledge many stimulating discussions with 
P.~R. King and R.~C. Ball.
One of the authors TBL acknowledges a studentship from B.P. Research.

\newpage
\begin{figure}
\caption{The evolution of a straight line at $t=0,4\times10^4,6\times10^4,
8\times10^4$ and $10^5$(bold line) iterations with 
$A=0.005$, $B=0.000012$ and $D=0.002$  via equation(5). 
}
\label{fig1}
\end{figure}

\begin{figure}
\caption{Configuration after 200,000 iterations (a) with oxbows and 
(b) without oxbow creation.
}
\label{fig2}
\end{figure}


\begin{references}
\bibitem[\dag]{Corres}
To whom all correspondance
should be addressed.
\end{references}
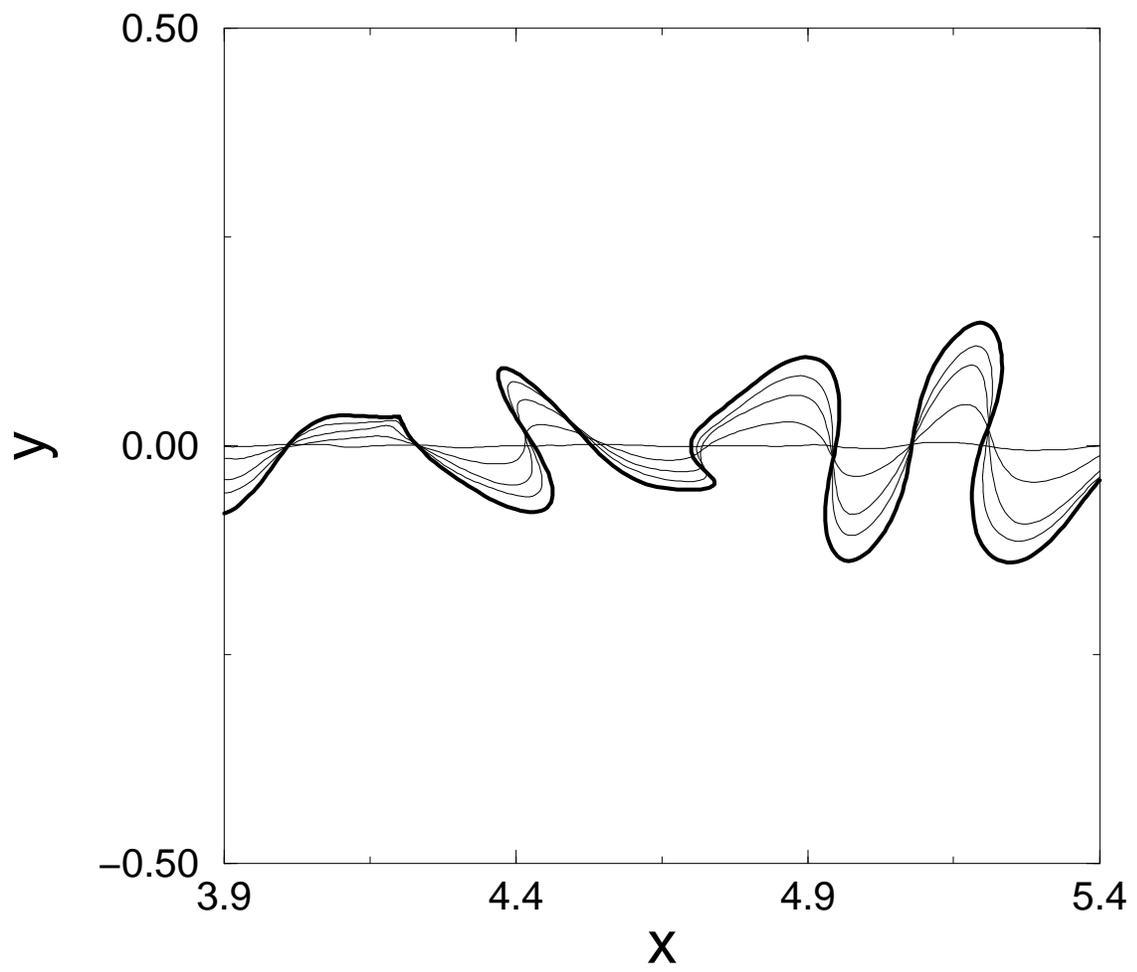
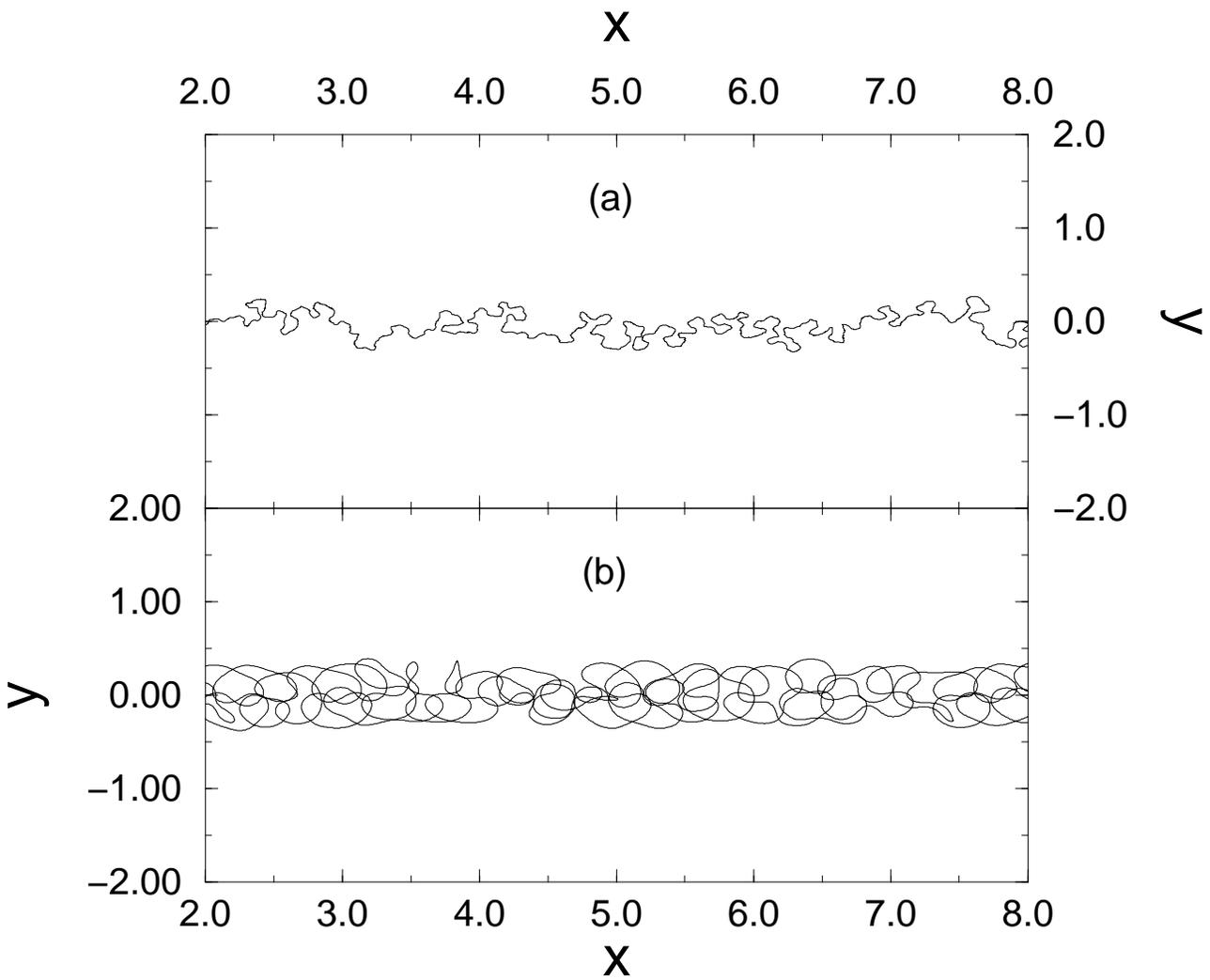
\end{document}